\documentclass[aps,prb,twocolumn,superscriptaddress,showpacs]{revtex4}
\usepackage{eurosym}
\usepackage{amsfonts}
\usepackage{amssymb}
\usepackage{amsmath}
\usepackage{graphicx}

\begin{document}

\title{Accounting for spin fluctuations beyond LSDA in the density
functional theory}
\author{L. Ortenzi}
\affiliation{Max-Planck-Institut f\"{u}r Festk\"{o}rperforschung, Heisenbergstra$\mathrm{%
\beta}$e 1, D-70569 Stuttgart, Germany}
\author{I.I. Mazin}
\affiliation{Naval Research Laboratory, 4555 Overlook Avenue SW, Washington, DC 20375, USA}
\author{P. Blaha}
\affiliation{TU Vienna, Institute of Materials Chemistry, Getreidemarkt 9/165-TC, A-1060
Vienna}
\author{L. Boeri}
\affiliation{Max-Planck-Institut f\"{u}r Festk\"{o}rperforschung, Heisenbergstra$\mathrm{%
\beta}$e 1, D-70569 Stuttgart, Germany}
\date{\today }

\begin{abstract}
We present a method to correct the magnetic properties of itinerant systems
in local spin density approximation (LSDA) and we apply it to the
ferromagnetic-paramagnetic transition under pressure in a typical itinerant
system, Ni$_{3}$Al. We obtain a scaling of the critical fluctuations as a
function of pressure equivalent to the one obtained within Moryia's theory.
Moreover we show that in this material the role of the bandstructure is
crucial in driving the transition. Finally we calculate the magnetic moment
as a function of pressure, and find that it gives a scaling of the Curie
temperature that is in good agreement with the experiment. The method can be
easily extended to the antiferromagnetic case and applied, for instance,
to the Fe-pnictides in order to correct the LSDA magnetic moment.
\end{abstract}

\maketitle


Density functional theory (DFT), in its most common implementations (local
spin density approximation, LSDA, with or without gradient corrections,
GGA), is in principle the only way to access the ground state of real
materials.~\cite{DFTreview} 
And indeed the agreement with experiment concerning the ground state
properties, such as crystal and electronic structures, is excellent,
especially for itinerant system, where local correlations play a minor role. 
Nevertheless, a well known problem of LSDA is the overestimation of the tendency
to magnetism in itinerant magnets near the quantum critical point (QCP).
This problem can be traced down to the fact that LSDA is essentially a mean
field theory, which does not take into account a detrimental effect of
near-critical fluctuations on the long-range magnetism. 
%
%
%
%
Thus, while the itinerant nature of systems like FeAl,~\cite{Petukhov}
Pd~\cite{mazin_pd} or the more recent and better known Fe-pnictides,~\cite%
{singh1,greenpaglione,liliaole} make LDA and GGA reproduce very well the
paramagnetic bandstructure, whenever a (magnetic) quantum critical point
(QCP) is approached the theory fails miserably. %
The importance of this problem is demonstrated by the amount of papers
dealing with the problem of correcting the magnetic moment of Fe-pnicitides.~%
\cite{lorenzana,negativeU,Philip,Silke,Haule,Gutzwiller} There, the usual
argument is that correlations beyond mean field suppress the (LSDA) local
ordered moment. It was shown that one can reduce the calculated magnetic
moment by using the LDA+U method with a negative U~\cite{negativeU}, but
there is no physical justification for this procedure. Such many-body
approaches as Dynamic Mean Field Theory (DMFT)~\cite{Philip,Silke,Haule} and
Gutzwiller~\cite{Gutzwiller} were also successfully used; since these
methods introduce additional fluctuations into the system, they obviously
work in the right direction. However, the concept of substituting long-range
critical fluctuations by the on-site ones is rather questionable.~\cite%
{Petukhov} Furthermore the effect of non-local fluctuations was recently
found to be crucial, also in localized models, whenever the critical
behavior is analyzed\cite{toschi}, and in any event computational load in
these methods is incomparably heavier than in LDA calculations.

For these reasons we propose a different approach which corrects LSDA within
DFT and takes into account the itinerant nature of the system. Our method is
easy to implement,~\cite{patch} carries no additional computational cost,
and has a transparent physical justification. 
The approach we describe in the following is based on the idea that unknown
more accurate DFT is very close to the conventional LSDA-GGA functional, but
the energy gain due to spin polarization (\textquotedblleft Stoner
interaction\textquotedblright , using the DFT parlance) is reduced by about
as much as the Moriya selfconsistent renormalization (SCR) theory~\cite%
{moriya1}, successfully used before~\cite{mazin_pd,Ni3AlNi3Ga,shimizu},
suggests. For this reason we call it reduced Stoner theory (RST). 
In fact we show that there is a one to one and well defined connection
between our method and the Moriya SCR theory in accounting for the effect of
spin fluctuations in itinerant magnets. This allows us to make an \emph{%
ab-initio} prediction of the magnetic moment as a function of pressure for
the archetypical Ni$_{3}$Al itinerant ferromagnet, with the correct scaling
of the Curie temperature, which until now was impossible. For Ni$_{3}$Al
indeed, the effect of spin fluctuations on physical properties both in the
magnetic state (magnetic moment and susceptibility) and in the paramagnetic
one (dc resistivity) was demonstrated both theoretically~\cite%
{Ni3AlNi3Ga,quantitativemoryia1} and experimentally.~\cite%
{Ni3Al-magnetic1,lonzarich} 

Applying the RST we estimate the pressure dependence of the contribution due
to spin fluctuations on the magnetic moment of Ni$_{3}$Al. Surprisingly we
find that this contribution is almost pressure independent. This means that
spin fluctuations only act in shifting the Stoner condition. The rest is
done by the bandstructure. As a consequence we show that the way in which
the spin fluctuations renormalize the effective Stoner interaction is
encoded in the bandstructure itself. This is generally true but in Ni$_{3}$%
Al it becomes particularly evident due to the perfect scaling of the density
of states (DOS) with pressure.

Having established a link between the RST and the SCR theory indeed, we are
able to give a reliable estimate of the Stoner parameter in itinerant
magnets. 
Moreover if one reverses the logic a comparison between the experiment and
the RST results gives an easy and reliable estimate of the spin
fluctuations acting in the system. In the following we present the bare LSDA
results for Ni$_{3}$Al under pressure and we interpret them within the
so-called extended Stoner Theory (EST).~\cite{extended,krasko,extended2} 
We explain that the overestimation of both the magnetic moment at zero 
pressure, and the critical pressure $P_{\text{c}}$ come essentially by the
overestimation of the Stoner parameter in LSDA.
After that we introduce the formalism for correcting the LSDA behavior and
we present the scaling equations used in the RST. In the last part of the
paper we apply these scaling equations to the ferromagnetic-paramagnetic
transition of Ni$_{3}$Al under pressure. 
\begin{figure}[t]
\centerline{\includegraphics*[angle=0,width=1.0
\columnwidth]{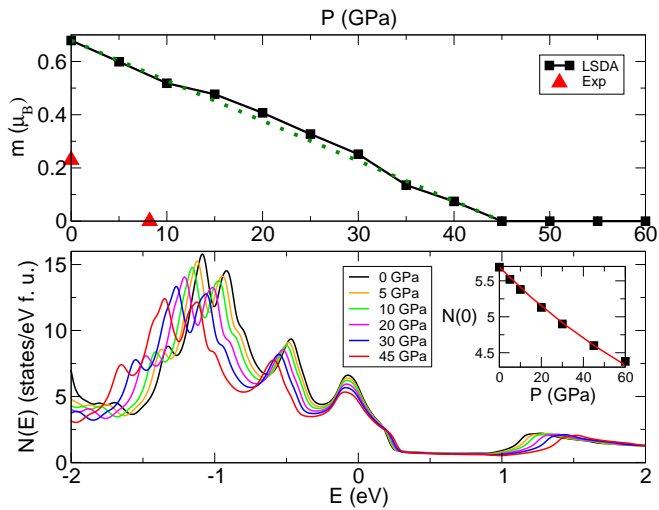}}
\caption{{\protect\footnotesize {Top: Magnetic moment of Ni$_{3}$Al per Ni
atom calculated in LSDA (black dotted line). Both the magnetic moment at
zero pressure $m(0)$ and the critical pressure P$_{\text{c}}$ are
overestimated with respect to the experimental ones (red triangles). The
green dots mark a linear interpolation of the data. Bottom: Paramagnetic
density of states as a function of energy calculated for different pressures
in LDA. Inset: Density of states at the Fermi level calculated as a function
of pressure (black squares). The data are fitted with eq.~\protect\ref%
{Pscale} (red continuous line). }}}
\label{fig:DOSdiP}
\end{figure}

First we discuss the bare LDA (and LSDA) results for the magnetic transition
in Ni$_{3}$Al under pressure. 
Ni$_{3}$Al crystallizes in the ideal cubic Cu$_{3}$Au $cP4$ structure. 
We calculated the equilibrium lattice parameter $a=3.4825$ \AA ~in
LDA~\cite{functional} and we found it to be  $\approx2\%$ smaller than the 
experimental one.~\cite{technical} The calculated magnetic moment 
at $P=0$ GPa is $m(0)=0.68$ $\mu_{\text{B}}$ in LSDA ($0.73$ $\mu _{\text{B}}$
in GGA) in reasonable agreement with previous results~\cite{Ni3AlNi3Ga}.
After that we calculated the magnetic moment as a function of pressure and, as 
shown in the top of Fig.~\ref{fig:DOSdiP}, we found that it decreases approximately linearly up to the critical pressure $P_{\text{c}}=45$ GPa, much larger than the experimental one, which is $P_{\text{c}}^{\text{Exp}}\approx 8$ GPa.~\cite{Ni3Al-magnetic1} This behavior can be easily understood within the EST.~%
\cite{extended,krasko,extended2} In fact, as shown in Fig.~\ref{fig:DOSdiP}
(bottom), the paramagnetic DOS $N(E)$ calculated in LDA scales almost
perfectly with the pressure $P$ as: 
\begin{equation}
N(E,P)=\frac{N(E*Z(P),0)}{Z(P)},  \label{Pscale}
\end{equation}%
in a wide energy range around the Fermi level taken as the origin, where $Z(P)=1+0.005\ast P$ with P measured in GPa.
This can be related to magnetization $via$ the EST~\cite%
{extended,krasko,extended2}, which combines the Stoner criterion with the
Andersen's force theorem to show that in the lowest order in the magnetic
moment $m$ the total LSDA energy is given in terms of the non magnetic DOS
per spin: 
\begin{equation}
E(m)=\frac{1}{2}\int_{0}^{m}\frac{m^{\prime }dm^{\prime }}{\tilde{N}%
(m^{\prime })}-\frac{1}{4}Im^{2}  \label{Edim}
\end{equation}%
where $\tilde{N}(m^{\prime })$ is the extended DOS defined in Ref.~%
\onlinecite{extended} as the average DOS over an energy interval equal to
the exchange splitting at a given $m,$ and $I$ is the Stoner parameter.
Stationary solutions appear where $\tilde{N}(m)=1/I,$ and $d\tilde{N}(m)/dm<0
$, forming a stable or a metastable (metamagnetic) ferromagnetic state with
the magnetic moment $m$. If $N(E)$ scales according to eq.~\ref{Pscale}, so
does also $\tilde{N}(m)$. Comparing eq.~\ref{Edim} with the fixed spin
moment calculation we found the Stoner parameter in LSDA to be $%
I^{LSDA}\approx 0.41$ eV independent on pressure (see appendix~\ref{appendix}%
). In GGA the Stoner parameter is 17\% larger, $I^{GGA}=1.17I^{LSDA}$. In
the following we will consider only the LSDA results. Around the minimum of $%
E(m)$ $\tilde{N}(m)$ decreases monotonically. Thus a reduction of $I$ shifts
the Stoner condition [$\tilde{N}(m)=1/I$] toward a smaller magnetic moment,
bringing the LSDA results in agreement with the experiment. The aim of this
paper is to find a way for reducing the LSDA Stoner parameter and getting in
this way the correct magnetic moment in itinerant systems. %
%
%
%
%
%
%
%
%
In order to gain some understanding of how the standard LSDA needs to be
corrected to account for spin fluctuations, let us recall the SCR theory for
critical ferromagnets. It starts with the Ginzburg-Landau expansion, 
\begin{equation}
E(M)=a_{0}+a_{2}M^{2}/2+a_{4}M^{4}/4+a_{6}M^{6}/6+\cdots ,  \label{EdiM}
\end{equation}%
where $M$ is the magnetic moment and $E$ is the LSDA energy. It is then
assumed that $M$ fluctuates around the average value $\bar{M},$ so
that $M=\bar{M}+\delta M$. Assuming that $\delta M$ follows a Gaussian
distribution such that $\left\langle \delta M^{2}\right\rangle =\xi ^{2},$
we can rewrite eq.~\ref{EdiM} in terms of $\bar{M}$ as 
\begin{equation}
E(\bar{M})=\tilde{a}_{0}+\tilde{a}_{2}\bar{M}^{2}/2+\tilde{a}_{4}\bar{M}%
^{4}/4+\tilde{a}_{6}\bar{M}^{6}/6+\cdots ,
\end{equation}%
where the explicit formula for the renormalized coefficients is given in
Ref.~\onlinecite{mazin_pd}, and the second order coefficient, $\tilde{a}%
_{2}=a_{2}+(5/3)a_{4}\xi ^{2}+(35/9)a_{6}\xi ^{4}+\cdots .$ If we restrict
our expansion by the second order, this SCR procedure is equivalent to
renormalizing spin susceptibility.

The amplitude of the spin fluctuations, $\xi ,$ in principle, can be
obtained by the fluctuation-dissipation theorem (FDT): 
\begin{equation}
\xi ^{2}=\frac{4\hbar }{\Omega }\int d\mathbf{q}\int \frac{d\omega }{%
2\pi }\frac{1}{2}Im\chi (\mathbf{q},\omega )  \label{FDT}
\end{equation}%
where $\Omega $ is the volume of the unit cell.~\cite{mazin_pd} $\chi(%
\mathbf{q},\omega) $ in this formula is often expressed in terms of the
noninteracting (Lindhard) susceptibility, which in turn is expanded to the
lowest order in $q$ and $\omega ,$ $\chi _{0}(\mathbf{q},\omega )=\chi _{0}(%
\mathbf{0},0)-aq^{2}+ib\omega /q$\cite{mazin_pd}. The coefficients $a$ and $b
$ can be written down as functions of the Fermi velocity, averaged over the
Fermi surface. However, in order to calculate $\xi $ using this expression
one needs to apply a cutoff $q_{c}$ which is not a well defined quantity.

Let us now point out that the uniform spin susceptibility in the LSDA can be
written in a particularly simple form, namely $\chi ^{-1}=\delta
^{2}E/\delta M^{2}\mid_{M=0} =a_{2}=\frac{1}{2}\left[\frac{1}{N_{\sigma
}(E_{F})}-I\right]$ (this can be considered as a rigorous definition of the
LSDA Stoner parameter $I)$, where $N_{\sigma }(E_{F})$ is the paramagnetic
DOS per spin. Comparing this expression with the one for $\tilde{a}_2$ we
see that the SCR procedure is equivalent to renormalization of $I$ according
to 
\begin{equation}
\tilde{I}=I-(10/3)a_{4}\xi ^{2},  \label{I}
\end{equation}
which can also be written as $\tilde{I}=sI,$ where $s<1$ (see also appendix~%
\ref{appendix}).

This is a justification of the recipe of using $I$ as an adjustable
parameter~\cite{CuBiSO}, often used in critical ferromagnets empirically. In
this sense, $\tilde{I}$ can be perceived as derivable from an unknown more
accurate DFT, in a specific material. One can also \textquotedblleft
reverse-engineer\textquotedblright\ such an improved functional, using the
standard von Bart-Hedin scaling, 
\begin{eqnarray}
E_{xc} &=&\int \varepsilon _{xc}(n,m)n(\mathbf{r)}d\mathbf{r}  \label{vBH} \\
\varepsilon _{xc}(n,\zeta ) &=&\varepsilon _{xc}^{P}(n)+f(\zeta )\Delta
\varepsilon _{xc}(n),\label{vBH2}
\end{eqnarray}%
where $\zeta (\mathbf{r)}=m(\mathbf{r)}/n(\mathbf{r)},$ $\varepsilon
_{xc}^{P}$ and $\Delta \varepsilon _{xc}(n)$ do not depend on $m,$ and $%
f(\zeta )$ is a known function, and $n=(n_{\uparrow }+n_{\downarrow }),$ $%
m=(n_{\uparrow }-n_{\downarrow })$. The response to magnetism is entirely
defined by the $\Delta \varepsilon _{xc}(n)$ functional, as the energy
difference between the fully polarized and unpolarized electron gas:%
\begin{align}
\frac{\partial \varepsilon _{xc}}{\partial n}& =\frac{\partial \varepsilon
_{xc}^{P}}{\partial n}+f(\zeta )\frac{\partial \Delta \varepsilon _{xc}(n)}{%
\partial n}-f^{\prime }(\zeta )\Delta \varepsilon _{xc}(n)\frac{\zeta }{n} \\
\frac{\partial \varepsilon _{xc}}{\partial m}& =f^{\prime }(\zeta )\Delta
\varepsilon _{xc}(n)\frac{1}{n}
\end{align}%
Note that the charge potential also acquires a term that disappears when $%
\zeta =0.$ In this sense, it is impossible in a physically meaningful way to
reverse-engineer $f$ and $\Delta \varepsilon _{xc}$ in such a way that $%
\partial \varepsilon _{xc}/\partial m$ be scaled by a constant factor $s,$
and $\partial \varepsilon _{xc}/\partial n$ would remain the same. Rather, a
natural way to weaken the magnetism in this formalism is to scale $\Delta
\varepsilon _{xc}(n)$ in eq.~\ref{vBH2}. Then we will have the following set
of scaled equations: 
\begin{align}
\varepsilon _{xc}(n,\zeta )& =\varepsilon _{xc}^{P}(n)+sf(\zeta )\Delta
\varepsilon _{xc}(n) \\
\frac{\partial \varepsilon _{xc}}{\partial n}& =v_{xc}^{P}+s[f(\zeta )\Delta
v_{xc}(n)-f^{\prime }(\zeta )\Delta \varepsilon _{xc}(n)\frac{\zeta }{n}] \\
\frac{\partial \varepsilon _{xc}}{\partial m}& =sf^{\prime }(\zeta )\Delta
\varepsilon _{xc}(n)\frac{1}{n}
\end{align}%
where the part in the square brackets is simply the additional charge
potential that appears because of spin polarization. It is easy to verify
that this functional produces an exchange-correlation potential scaled by $s,
$ and the charge potential unchanged:%
\begin{eqnarray}
\tilde{V}_{\uparrow }(\mathbf{r)-}\tilde{V}_{\downarrow }(\mathbf{r)} &%
\mathbf{=}&s\mathbf{[}V_{\uparrow }(\mathbf{r)-}V_{\downarrow }(\mathbf{r)]}
\label{scale} \\
\tilde{V}_{\uparrow }(\mathbf{r)+}\tilde{V}_{\downarrow }(\mathbf{r)} &%
\mathbf{=}&\mathbf{[}V_{\uparrow }(\mathbf{r)+}V_{\downarrow }(\mathbf{r)]} 
\notag
\end{eqnarray}%
and the Stoner kernel $\delta ^{2}E_{xc}/\delta m(\mathbf{r)}^{2}$ also
scaled by $s,$ as we wanted. Eq.~\ref{scale} can be easily implemented~%
\cite{patch} and can be used to obtain correct magnetic moments and the
corresponding electronic structure in the materials near ferro \textit{or
antiferro}magnetic QCP. Moreover, given eq.~\ref{FDT} and~\ref{I} $s$ gives
also an indication of the strength of spin fluctuations acting in the
system. Below we use eq.~\ref{scale} for correcting the LSDA results
obtained in the previous section in the case of Ni$_{3}$Al itinerant
ferromagnet under pressure. %
%
As shown previously, Ni$_{3}$Al becomes paramagnetic under pressure~\cite%
{Ni3Al-magnetic1} and LSDA overestimates not only the magnetic moment, but
also the critical pressure $P_{\text{c}}$. Empirically, by comparing the
LSDA value of the magnetic moment with the experimental one for $P=0$ GPa
and for $P=6$ GPa, we found that in both cases the value of $s$ needed to
reconcile the LSDA result with experiment, using the scaling introduced in
eq.~\ref{scale}, is $s\approx 0.88$. This implies that $\xi $ is almost
pressure-independent between 0 and 6 GPa. This value of $s$ gives for $%
\tilde{I}$ the same value obtained by renormalizing $I$ within the EST (see
appendix~\ref{appendix}). 
In general, one expects that spin fluctuations become stronger closer to the
critical pressure, so that their average amplitude $\xi $ becomes larger,
and the scaling parameter $s=1-(10/3)a_{4}\xi ^{2}/I$ in eq.~\ref{I} smaller%
%
. On the other hand, $\xi $ is defined by averaging over the entire
Brillouin zone, and the fact that susceptibility at one particular point $%
\mathbf{q=0}$ diverges may or may not strongly affect $\xi $. 
In order to understand this result, we compared our calculations with Ref.~%
\onlinecite{Ni3AlNi3Ga}, where $\xi $ is calculated for $P=0$ GPa in the
approximations described above. By means of scaling arguments described
below, we found that the transition is driven entirely by the change in DOS
given by eq.~\ref{Pscale} while the renormalized Stoner parameter $\tilde{I}$
is, in the first approximation, pressure independent. 
%
Let us try to rationalize this fact, using the fluctuation-dissipation
theorem (eq. \ref{FDT}).

\begin{figure}[t]
\centerline{\includegraphics*[angle=0,width=1.0
\columnwidth]{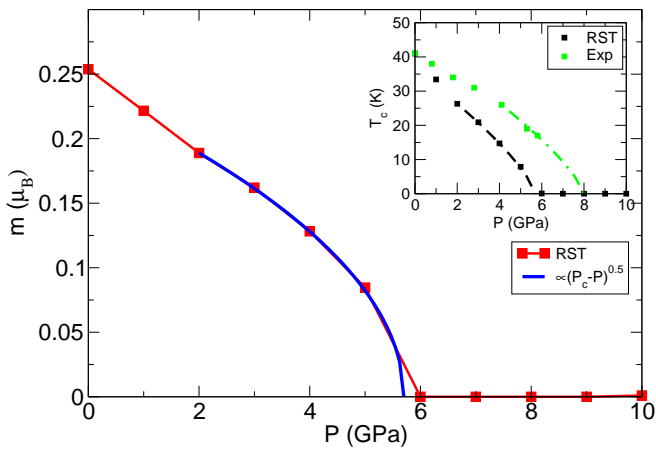}}
\caption{{\protect\footnotesize {Predicted magnetic moment as a function of
pressure calculated within the RST (red dotted line). The critical behavior $%
m\propto (P_{\text{c}}-P)^{1/2}$ (blue line) is followed between 2 and 6
GPa. The Curie temperature $T_{\text{c}}\propto m^{3/2}$ is shown in the
inset by black dots. Apart from the underestimation of the critical
pressure, the agreement with the experimental data taken from Ref.~
\onlinecite{Ni3Al-magnetic1} (green dots) is very good. In fact, both
theoretically and experimentally, the extrapolation of the data the data at
5.7 GPa (black dashed line) and 8 GPa (green dashed line) respectively, give
both the same critical behavior $T_{\text{c}}\propto (P-P_{\text{c}})^{3/4}$%
, as observed by Nicklovic \emph{et al.}~\protect\cite{Ni3Al-magnetic1} }}}
\label{mvsP}
\end{figure}

Given eq.~\ref{Pscale}, and since the Fermi velocity scales inversely with
the DOS, $v(P)=v(0)Z(P)$, in the approximation described in the beginning,
and used in Refs.~\onlinecite{Ni3AlNi3Ga}, we found that $\xi $ does
actually scale with pressure as $\xi (P)\propto \sqrt{\frac{1}{\Omega (P)}}$
where $\Omega (P)$ is the unit cell volume at pressure $P$.~\cite{footnote}
We found indeed that it does not diverge or even grow substantially near the
critical pressure (even though the model susceptibility at $q=0,$ $\omega =0$
does diverge there, $\tilde{I}N(E_{F})=1$). This result is particularly
important because it tells us that the effect of spin fluctuations in Ni$_{3}
$Al under pressure can be entirely accounted for by renormalizing the Stoner
factor (the rest is a bandstructure effect). Furthermore, the fact that the
correlation length associated with the fluctuations ($\xi $) does not
diverge approaching the transition, is compatible with the fact that the
system does not show other instabilities (like triplet superconductivity) at 
$P=P_{\text{c}}$. In fact if $\xi $ would go to infinity, another
(competing) instability could profit of this kind of long range correlation
in order to build up a competing order parameter. In fact, $\sqrt{\Omega (6%
\text{ GPa})/\Omega (0\text{ GPa})}=0.98,$ which implies that $\xi $ changes
only by 2\%. Applying the SCR theory starting from the fixed spin moment
calculations and using $\xi $ as a parameter (see appendix~\ref{appendix}),
we found that the best agreement with the experiment was achieved if $\xi $
changes slightly more, by 5\%, but simply using a pressure-independent $\xi $%
, corresponding to the scaling parameter $s=0.88$, provides, apart from some
underestimation of $P_{\text{c}}$, a very reasonable agreement with the
experiment. This choice of $s$ allows us also to make predictions about the
magnetic moment between $0$ and $P_{\text{c}},$ as shown in Fig.~\ref{mvsP}. 
%
%
%
%
%
%
%
%

There are no experimental measurements of magnetization as a function of
pressure between $0$ and $8$ GPa in Ni$_{3}$Al, but  the pressure dependence
of the Curie temperature $T_{\text{c}}$ was measured by Nicklovic \emph{et
al.}~\cite{Ni3Al-magnetic1} They also analyzed the data using a Landau
functional for the field%
\begin{equation}
\mathbf{H}=a_{2}\mathbf{m}+a_{4}\mathbf{m}^{3}-c\nabla ^{2}\mathbf{m}.
\label{Landaumagn}
\end{equation}%
By assuming $a_{2}\propto (P-P_{\text{c}})$ they found $T_{\text{c}}\propto
(P-P_{\text{c}})^{3/4}$ in good agreement with the experimental data. 
Within Moriya's theory~\cite{quantitativemoryia1} 
$T_{\text{c}}\propto m^{3/2}$ 
where $m=\mid \mathbf{m}\mid $. Assuming $a_{2}\propto (P-P_{\text{c}})$ in
eq.~\ref{Landaumagn} gives $m\propto (P-P_{\text{c}})^{\beta }$ with $\beta
=1/2$, from which  the result of Nicklovic \emph{et al.} follows.
%
%
%
%
%
%
%
In the following we show that the linearity of $a_{2}$ with respect to $%
(P-P_{\text{c}})$ is a consequence of eq.~\ref{Pscale}. For small values of
the magnetic moment, in the fluctuation-corrected LSDA described above, 
$a_{2}=\frac{1}{2N_{\sigma }(E_{F})}-\frac{\tilde{I}}{2},$ 
$c=0$%
%
%
%
%
, and $\tilde{I}$ is adjusted so as to have $N_{\sigma }(E_{F})\tilde{I}=1$
at $P=P_{c}$. Given eq.~\ref{Pscale} then $a_{2}=\frac{Z(P)/Z(P_{c})}{%
2N_{\sigma }(E_{F}),P_{c})}-\frac{\tilde{I}}{2}=\frac{\tilde{I}}{2Z(P_{c})}%
[\alpha (P-P_{c})]\propto (P-P_{\text{c}}),$ thus providing a microscopical
justification for the model of Ref.~\onlinecite{Ni3Al-magnetic1}. %
As shown in Fig.~\ref{fig:DOSdiP}, in LSDA (for large values of the magnetic
moment) we found $\beta \approx 1,$ while $\beta =0.5$ is recovered for
small values (see Fig~\ref{mvsP}). This is due to 
the fact that at large magnetic moment the coefficient $a_{2}$ must be
corrected by adding high order terms. %
%
%
%
Finally, using Moriya's relation for $T_{\text{c}}$ we find $T_{\text{c}%
}\propto (P-P_{\text{c}})^{3/4}$ in full agreement with Nicklovic \emph{et
al.} %
%
%
The disagreement with experiment in Fig.~\ref{mvsP} (inset) concerns only
the underestimation of $P_{\text{c}}$ caused by: the approximation of $s$ as a
constant value and the underestimation of the equilibrium lattice parameter.~\cite{technical} %
%
%
%

To summarize, in this paper we have described a simplified method for
accounting for near-critical spin-fluctuations within the DFT. The method
amounts to scaling the DFT exchange-correlation field by a phenomenological
constant $s$, and subsequent self-consistent solution of the Kohn-Sham
equations. This phenomenological constant can also be, in principle,
calculated via the fluctuation dissipation theorem~\cite{mazin_pd}, and in
this sense is equivalent to the SCR theory by Moriya. Our method is
complementary to the widely used similarly semi-phenomenological LDA+U, and
plays for itinerant near-critical magnets the same role as LDA+U for systems
near a Mott-Hubbard transition. We apply this method to the ferromagnetic
QCP in Ni$_{3}$Al under pressure. We show that, due to a particular scaling
property of the bandstructure, the parameter $s$ is constant with pressure.
In this way the method becomes completely \emph{ab initio}. In fact, the
ferromagnetic-paramagetic transition in Ni$_{3}$Al is driven by the band
structure changes under pressure, while the feedback to the critical
fluctuations (parameterized by $s$) is small. Using this formalism, we make
a prediction of the magnetic moment as a function of pressure, which
recovers the critical exponent for the magnetization $\beta =1/2$ for small
magnetic moments and explains the observed dependence of the Curie
temperature on pressure. The new method should be useful in cases when one
needs to calculate electronic properties of materials where LSDA
overestimates the tendency to magnetism, or when one wants to monitor
theoretically the evolution of the electronic structure from non magnetic to
magnetic and get an estimate of the spin fluctuation amplitude.

This research was supported by the Deutsche Forschungsgemeinschaft under
Priority Program 1458, grant number Boe/3536-1. We are grateful to Alaska
Subedi, Tobias Schickling and Mark Hoeppner for a careful reading of our 
manuscript.

\appendix

\section{Connection between RST and SRC via EST}

\label{appendix} 
\begin{figure}[!t]
\begin{center}
\centerline{\includegraphics*[angle=270,width=1.0
\columnwidth]{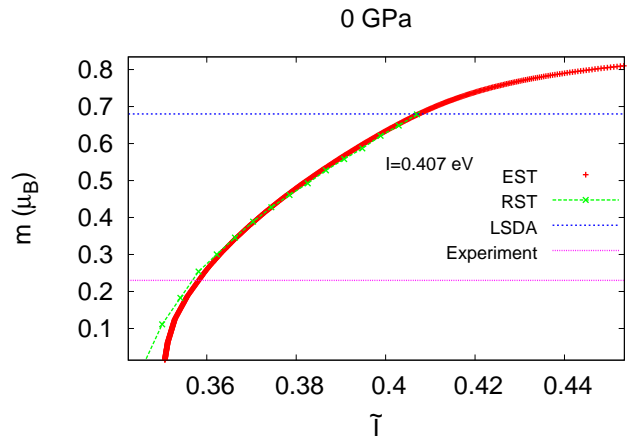}}
\end{center}
\caption{{\protect\footnotesize {Magnetic moment $m$ as a function of the
reduced Stoner parameter $\tilde{I}=s\ast I$ for Ni$_{3}$Al at $0$ GPa in
EST (red dots) and RST (green dotted line). The bare value of te Stoner
parameter $I=0.407$ eV was chosen in order to to have the same value of $m$
for $s=1$.}}}
\label{fig:mvsI0}
\end{figure}

If we fit the fixed spin moment calculations with a Landau
expansion up to the 8-th order we find the same result as in Ref.~%
\onlinecite{Ni3AlNi3Ga}. However, as opposed to that work, we do not want to
apply explicitly a fluctuation-induced renormalization, but renormalize $I$
directly, according to eq.~\ref{scale}, that is, to scale down \emph{%
ab-initio} the Stoner parameter in order to obtain the correct magnetic
moment for Ni$_{3}$Al.
To make a direct connection between the Landau approach and the RST
approach, we can use the EST. Indeed, as discussed above, the RST method is
equivalent to the SCR method, if only the second order coefficient $a_{2}$
(the inverse magnetic susceptibility) is renormalized. We can expand the EST
total energy in eq.~\ref{Edim} in $m,$ and then renormalizing the Stoner
parameter becomes exactly equivalent to a renormalization of $a_{2}$.
\begin{figure}[!t]
\begin{center}
\centerline{\includegraphics*[angle=270,width=1.0
\columnwidth]{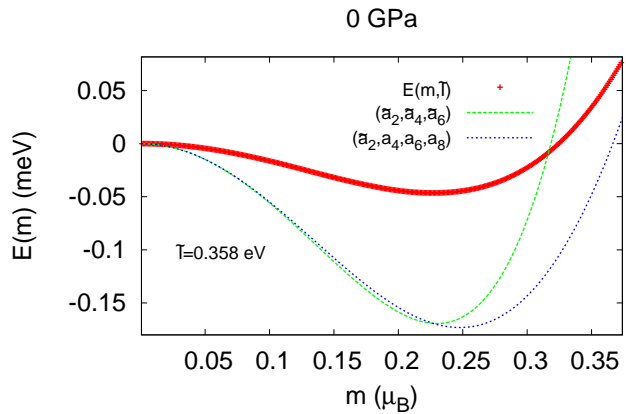}}
\end{center}
\caption{{\protect\footnotesize {Energy $E$ as a function of the magnetic
moment $m$ for $P=0$ GPa. The red dots show the curve obtained by means of
EST with a renormalized Stoner parameter $\tilde{I}=0.358$. The green and
the blue lines mark respectively the Landau functional where all the
coefficients are renormalized and the one where only the first coefficient
of the expansion is renormalized.}}}
\label{fig:comparison0GPa}
\end{figure}%
Fig.~\ref{fig:comparison0GPa} shows that, in fact, only the $a_{2}$
renormalization affects the equilibrium magnetic moment, while the
renormalization of the higher order coefficients influences $E(m)$ at $m$
larger than the experimental magnetic moment. Note that the red dotted curve
(EST) in Fig.~\ref{fig:comparison0GPa} corresponds to rigid splitting of the
unpolarized bands\cite{extended}, while the coefficients of the Landau
expansion were obtained by fitting fully selfconsistent magnetic
calculations.~\cite{Ni3AlNi3Ga}
In Fig.~\ref{fig:mvsI0}, is shown the equilibrium magnetic moment $m$ as a
function of the reduced Stoner parameter $\tilde{I}=sI,$ obtained both in
the EST and by self-consistent calculations using the exchange scaling $s$
of eq.~\ref{scale}, at $P=0$ GPa. The value of $I$ is chosen, in the former
case, in such a way that the LSDA and EST magnetic moments have the same
value for $s=1$. 
The agreement between the two curves is perfect and the value of $\tilde{I}$
found in this way is in perfect agreement with the one extracted from the
comparison of the EST with the renormalized Landau expansion in Fig.~\ref%
{fig:comparison0GPa}.


\begin{thebibliography}{999}
%

\bibitem{DFTreview} R. O. Jones and O. Gunnarsson Rev. Mod. Phys. \textbf{61}%
, 689 (1989). 


\bibitem{Petukhov} A. G. Petukhov, I. I. Mazin, L. Chioncel and A. I.
Lichtenstein Phys. Rev. B \textbf{67}, 153106 (2003) 


\bibitem{mazin_pd} P. Larson, I.I. Mazin and D. J. Singh, Phys. Rev. B 
\textbf{69}, 064429 (2004) 


\bibitem{singh1} D. J. Singh, Physica C \textbf{469}, 418,424 (2009).

\bibitem{greenpaglione} J. Paglione and R. L. Greene, Nat. Phys. \textbf{6},
645 (2010).

\bibitem{liliaole} O. K. Andersen, and L. Boeri Ann.Phys. \textbf{523}, 8-50
(2011). 


\bibitem{lorenzana} G. Giovannetti, Carmine Ortix, M. Marsman, M. Capone, J.
van den Brink, and Jos\'{e} Lorenzana Nat. Comm. \textbf{2}, 398 (2011). 


\bibitem{negativeU} J. Ferber, Y.-Z. Zhang, H. O. Jeschke, and R. Valent$%
\acute{\imath}$, Phys. Rev. B \textbf{82}, 165102 (2010) 

\bibitem{Philip} P. Hansmann, R. Arita, A. Toschi, S. Sakai, G. Sangiovanni,
and K. Held Phys. Rev. Lett. \textbf{104}, 197002 (2010) 


\bibitem{Silke} Philipp Werner, Michele Casula, Takashi Miyake, Ferdi
Aryasetiawan, Andrew J. Millis, Silke Biermann, Nat. Phys. {\bf 8}, 333 (2012).



\bibitem{Haule} Z. P. Yin, K. Haule, and G. Kotliar, Nat. Mat. \textbf{10},
932 (2011) 


\bibitem{Gutzwiller} T. Schickling, F. Gebhard, J. B\"unemann,
L. Boeri, O. K. Andersen, and W. Weber, Phys. Rev. Lett. \textbf{108}%
, 036406 (2012) 

\bibitem{toschi} G. Rohringer, A. Toschi, A. Katanin and K. Held Phys. Rev.
Lett. 107, 256402 (2011).

\bibitem{patch} A patch for the popular WIEN2k code is available by request
from P. Blaha %

\bibitem{moriya1} T. Moriya, \emph{Spin Fluctuations in Itinerant Electron
Magnetism} (Springer, Berlin, 1985). 


\bibitem{Ni3AlNi3Ga} A. Aguayo, I. I. Mazin and D. J. Singh, Phys. Rev.
Lett. \textbf{92}, 147201 (2004). 


\bibitem{shimizu} M. Shimizu, Rep. Prog. Phys. \textbf{44}, 329 (1981). 


\bibitem{quantitativemoryia1} Y. Takahashi, T. Moriya, J. Phys. Soc. Jpn. 
\textbf{54}, 1592 (1985) and references therein. 



\bibitem{lonzarich} G. G. Lonzarich and L. Taillefer, J. Phys. C: Solid
State Phys. \textbf{18}, 4339 (1985). 



\bibitem{Ni3Al-magnetic1} P. G. Niklowitz, F. Beckers and G. G. Lonzarich,
G. Knebel, B. Salce, J. Thomasson, N. Bernhoeft, D. Braithwaite, and J.
Flouquet, Phys. Rev. B \textbf{72}, 024424 (2005). 



\bibitem{extended} O.K. Andersen, J. Madsen, U.K. Poulsen, O. Jepsen, and J.
Koll$\acute{\text{a}}$r Physica B\&C \textbf{86-88}, 249 (1977). 


\bibitem{krasko} G. L. Krasko,
 \textbf{36}, 8565 (1987). 


\bibitem{extended2} I. I. Mazin, D. J. Singh Phys. Rev.B \textbf{56}, 2556
(1997). 











\bibitem{functional} J.P. Perdew and Y. Wang, Phys.Rev. B \textbf{45}, 13244
(1992).

\bibitem{technical} Our LDA and LSDA calculations were done using the
general potential linearized augmented-plane-wave (LAPW) method as
implemented in the \textsc{wien2k} package.~\cite{wien2k,functional} Up to
1330 $\mathbf{k}$ points were used in the self-consistent calculations with
an LAPW basis defined by the cutoff $R_{S}K_{\text{max}}=9$ both in the
magnetic and in the non magnetic calculations. A larger number of 4960 $%
\mathbf{k}$ points were used for calculating the non magnetic DOS. 
We fitted the total energy as a function of the unit cell volume with the
equation of state that we used is the one by Murnaghan:~\cite{EOS} $E=E0+[B%
\frac{V}{B^{\prime }}(\frac{(V0/V)^{B^{\prime }}}{(B^{\prime }-1)}+1)-\frac{%
B\ast V0}{(B^{\prime }-1)}]/14703.6$ where $V/V_{0}$ is the volume
compression. $P=B/B^{\prime B^{\prime }}-1)$ with $V_{0}=284.5961$ $Bohr^{3}$%
, $B=237.0284$ GPa, $B^{\prime }=3.8413$, $E_{0}=-9594.376522$ $Ry$. $B$ and 
$B^{\prime }$ are the Bulk modulus and derivative. The lattice parameters
both in the magnetic and in the non magnetic case were found to be the same. 




\bibitem{CuBiSO} L. Ortenzi, S. Biermann, O. K. Andersen, I.I. Mazin, L.
Boeri, Phys. Rev. B \textbf{83}, 100505(R) (2011). 















\bibitem{EOS} Murnaghan F.D., Proc. Natl. Acad.Sci. USA \textbf{30}, 244
(1944).








































\bibitem{wien2k} http://www.wien2k.at.

\bibitem{LSDA} J.P. Perdew and Y. Wang, Phys.Rev. B \textbf{45}, 13244
(1992).

\bibitem{DFT:EOS} S. K. Saxena, J. Phys. Chem. Sol. \textbf{65}, 1561 (2004).

\bibitem{footnote} 
In Refs.~\onlinecite{mazin_pd,Ni3AlNi3Ga}, $\xi $ is defined as 
\begin{equation*}
\xi ^{2}(P)=\frac{b(P)v_{\text{F}}^{2}(P)N^{2}(E_{\text{F}},P)}{%
8a^{2}(P)\Omega (P)}[Q^{4}\log (1+Q^{-4})+\log (1+Q^{4})]
\end{equation*}%
where: $a(P)=\frac{1}{12}\frac{d^{2}\langle N(E_{\text{F}},P)v_{x}^{2}(P)%
\rangle }{dE_{\text{F}}^{2}},$\newline
and $b(P)=\frac{1}{2}\langle N(E_{\text{F}},P)v^{-1}(P).$\newline
Since the Fermi velocity scales inversely with the DOS, $v(P)=v(0)Z(P),%
$ these parameters scale as:\newline
$a(P)=a(0)Z(P),$ and $b(P)=b(0)/Z^{2}(P).$\newline
The last parameter, $Q,$ depends on the cutoff vector, $Q=q_{c}\sqrt{%
a(P)/b(P)v_{\text{F}}(P)}$. A possible choice for $q_{c}$ is $q_{c}=\sqrt{%
N(E_{F})/a},$ because at that point the expansion of $\chi $ loses its
physical meaning ($\chi _{0}$ changes sign). Then $Q=\sqrt{N(E_{F},P)/b(P)v_{%
\text{F}}(P)}$ does not depend on pressure. This implies that $\xi $ does
actually scale with pressure as $\xi (P)\propto \sqrt{\frac{1}{\Omega (P)}}$ 
\end{thebibliography}
\end{document}